\documentclass{article}

\usepackage{epsfig} 
\usepackage{amsmath}
\usepackage{amsfonts}
\usepackage{graphicx}

\date{\today}

\newcommand{\insertplot}[5]{\begin{figure}
 \hfill\hbox to 0.05in{\vbox to #5in{\vfill
 \inputplot{#1}{#4}{#5}}\hfill}
 \hfill\vspace{-.1in}
 \caption{#2}\label{#3}
 \end{figure}}
 \newcommand{\inputplot}[3]{
 \special{ps: plotfile #1}
}
\newcounter{fig}

\newcommand{\beq}{\begin{equation}}
\newcommand{\eeq}{\end{equation}}
\newcommand{\beqs}{\begin{eqnarray}}
\newcommand{\eeqs}{\end{eqnarray}}

\numberwithin{equation}{section}
\newcommand{\be}{\begin{equation}}
\newcommand{\ee}{\end{equation}}
\newcommand{\bea}{\begin{eqnarray}}
\newcommand{\eea}{\end{eqnarray}}

\begin{document}

\title{$d\geq 5$  static black holes with $S^2\times S^{d-4}$ 
 \\event horizon topology} 
  
 \author{
 {\large Burkhard Kleihaus}, {\large Jutta Kunz} 
and {\large Eugen Radu}  
\\ 
\\
{\small Institut f\"ur Physik, Universit\"at Oldenburg, Postfach 2503
D-26111 Oldenburg, Germany}
}
\maketitle 
 
\begin{abstract}
We present numerical evidence for the existence of new black hole solutions
in $d\geq 6$ spacetime dimensions.
They approach asymptotically the Minkowski background and 
have an event horizon topology
$S^2\times S^{d-4}$.
These static solutions share the basic properties
of the nonrotating black rings in five dimensions,
in particular the presence of a conical singularity.
\end{abstract}

\section{ Introduction}
 Perhaps the most interesting development of the last decade in black hole physics
was Emparan and Reall's discovery  
of the black ring solution 
in $d=5$ spacetime dimensions \cite{Emparan:2001wn,Emparan:2001wk}. 
This asymptotically flat solution of the Einstein equations  has a horizon with topology $S^2\times S^1$,
while the Myers-Perry black hole \cite{Myers:1986un} has a  horizon topology $S^3$.
This made clear that a number of well known results in $d=4$ gravity
do not have a simple extension  to higher dimensions. 

The rapid progress following the discovery in \cite{Emparan:2001wn,Emparan:2001wk}
provided a rather extensive picture of the solutions' 
landscape for the five dimensional case \cite{Emparan:2006mm}, including configurations with abelian matter fields
\cite{Elvang:2003yy}-\cite{Chng:2008sr}.
Physically interesting solutions describing superposed black objects (black saturns \cite{Elvang:2007rd}, 
bycicling black rings \cite{Elvang:2007hs},
di-rings \cite{Iguchi:2007is}, concentric rings \cite{Gauntlett:2004wh}) were also constructed.
In all cases, the rotation provides the centrifugal repulsion that allows a regular solution to exist.
In the static limit, all known $d=5$ asymptotically flat
solutions with a nonspherical horizon topology possess
a conical singularity or other pathologies  \cite{Emparan:2001wk},
\cite{Kunduri:2004da}, \cite{Chng:2008sr}.

Despite the presence of several partial results in the literature,
the $d>5$ case remains largely unexplored (for a review, see \cite{Obers:2008pj}).
Moreover, it becomes clear that as the dimension increases, the phase structure
of the solutions
becomes increasingly intricate and diverse.
The main obstacle stopping the progress in this field seems to be the absence of
closed form solutions (apart from the Myers-Perry black holes), which were very useful in $d=5$.
Although the discovery in \cite{Emparan:2001wn,Emparan:2001wk}
 was based on '{\it educated guesswork}', (and knowledge of the 
four dimensional C-metric),
several systematic methods have been developed to  construct
$d=5$ black ring solutions and their generalizations.
Since no general framework seems to exist for $d>5$,
the issue of constructing black objects
with a nonspherical horizon topology 
was considered by using approximations,
in particular the method of asymptotic expansions \cite{Emparan:2007wm},\cite{Emparan:2009cs}.
The central assumption here is that some black objects
can be approximated by a certain very thin
black brane curved into a given shape.
However, this method has limitations; for
example it is supposed to work only if the length scales involved are
widely separated.

Perhaps a more straightforward approach to this problem would be 
to find such configurations numerically, as solutions of partial differential
equations with suitable boundary conditions. 
This paper aims at a first step in this direction,
since we propose a framework for a special class of solutions.
As the simplest example of
$d>5$ black objects with a nonstandard topology of the event horizon,
we present 
numerical evidence  for the existence of 
static solutions with a $S^2\times S^{d-4}$
topology of the event horizon in $d=6$ and $7$ dimensions.
These solutions share most of the features of
the $d=5$ static black ring, which is known in closed form \cite{Emparan:2001wk}.
In particular, they possess a conical singularity and approach the 
Schwarzschild-Tangherlini metric in a certain limit.

\section{A general ansatz and particular solutions}
\subsection{The equations}
We consider the following metric ansatz:
\begin{eqnarray}
\label{metric} 
ds^2=-f_0(r,z)dt^2+f_1(r,z)(dr^2+dz^2)+f_2(r,z)d\psi^2+f_3(r,z)d\Omega_{d-4}^2,
 \end{eqnarray}
where $d\Omega^2_{d-4}$ is the unit metric on $S^{d-4}$ and $0\leq r<\infty,$ $-\infty< z<\infty$ and
$\psi$ is an angular coordinate, with $0\leq \psi \leq 2 \pi$ in the asymptotic region.
Thus these coordinates have a rectangular boundary and  are suitable for numerics.

An appropriate combination 
of the Einstein equations,
 $G_t^t=0,~G_r^r+G_z^z=0$, $G_{\psi}^{\psi}=0$,
 and  $G_{\Omega}^{\Omega}=0$ (with $G_\mu^\nu$ the Einstein tensor),
yields the following set of equations for the functions $f_1,~f_2,~f_3$ and $f_0$ 
\begin{eqnarray}
\nonumber
&&\nabla^2 f_0-\frac{1}{2f_0}(\nabla f_0)^2+\frac{1}{2f_2}(\nabla f_0)\cdot( \nabla f_2)+\frac{(d-4)}{2f_3}(\nabla f_0)\cdot( \nabla f_3)=0,
\\
\nonumber
&&\nabla^2 f_1-\frac{1}{f_1}(\nabla f_1)^2-(d-4)(d-5)\frac{f_1}{4f_3^2}(\nabla f_3)^2-\frac{f_1}{2f_0f_2}(\nabla f_0)\cdot( \nabla f_2)
\\
\nonumber
&&{~~~~~~~~}- \frac{(d-4)f_1}{2f_0f_3}(\nabla f_0)\cdot( \nabla f_3)- \frac{(d-4)f_1}{2f_2f_3}(\nabla f_2)\cdot( \nabla f_3)+\frac{(d-4)(d-5)f_1^2}{ f_3}=0,
\\
\label{eqs1} 
&&\nabla^2 f_2-\frac{1}{2f_2}(\nabla f_2)^2+\frac{1}{2f_0}(\nabla f_0)\cdot( \nabla f_2)+\frac{(d-4)}{2f_3}(\nabla f_2)\cdot( \nabla f_3)=0,
\\
\nonumber
&&\nabla^2 f_3+\frac{(d-6)}{2f_3}(\nabla f_3)^2+\frac{1}{2f_0}(\nabla f_0)\cdot( \nabla f_3)
+\frac{1}{2f_2}(\nabla f_2)\cdot( \nabla f_3)-2(d-5)f_1=0,
\end{eqnarray}
where we define 
$
(\nabla U) \cdot (\nabla V)=\partial_r U \partial_r V+ \partial_z U \partial_z V,~~~
\nabla^2 U=\partial_r^2U+\partial_z^2 U.
$

The remaining Einstein equations $G_z^r=0,~G_r^r-G_z^z=0$
yield two constraints. Following \cite{Wiseman:2002zc}, we note that
setting $G^t_t=G^{\varphi}_{\varphi} =G^r_r+G^z_z=0$
in $\nabla_\mu G^{\mu r}=0$ and $\nabla_\mu G^{\mu z}=0$, we obtain the Cauchy-Riemann relations
\begin{eqnarray}
\partial_z\left(\sqrt{-g} G^r_z \right) +
  \partial_r\left(  \sqrt{-g} \frac{1}{2}(G^r_r-G^z_z) \right)
= 0 ,~~
 \partial_r\left(\sqrt{-g} G^r_z \right)
-\partial_z\left(  \sqrt{-g} \frac{1}{2}(G^r_r-G^z_z) \right)
~= 0 . 
\end{eqnarray}
Thus the weighted constraints satisfy Laplace equations,
and the constraints are fulfilled,
when one of them is satisfied on the boundary 
and the other at a single point
\cite{Wiseman:2002zc}. 
\subsection{$d=5$}
For $d=5$, this is the standard coordinate system used to study static, axisymmetric
solutions of the vacuum Einstein equations, in which case the sphere $\Omega_{d-4}$ reduces to a single angular coordinate $\varphi$
(with $0 \leq \varphi \leq 2 \pi$).
The above equations present in this case a variety of 
physically interesting solutions 
which can be uniquely characterized
by the boundary conditions on the $z-$axis, known as the {\it rod-structure}.
In this approach, the $z-$axis is divided into $N$ intervals (called rods of the 
solution), $[-\infty, z_1]$,
$[z_1,z_2]$,$\dots$, $[z_{N-1},\infty]$.
A necessary condition for a regular solution is that only one of the functions
$f_0(0,z)$, $f_2(0,z)$, $f_3(0,z)$ becomes zero for a given rod,
except for isolated points between the intervals.
 For the static case discussed here, a horizon corresponds to 
 a timelike rod where $f_0(0,z)=0$ while $\lim_{r\to 0}f_0(r,z)/r^2>0$.
 There are also spacelike rods corresponding to compact directions  specified by the
 conditions $f_{a}(0,z)=0$, $\lim_{r\to 0}f_{a}(r,z)/r^2>0$, with $a=2,3$.
A semi-infinite spacelike rod corresponds to an axis of rotation, the associated coordinate being
a rotation angle.

One of the main advantages of this approach is that
the topology of the horizon is automatically imposed by the rod structure.
For example, the rod structure of a $d=5$ static  black ring solution  
consists of  
a semi-infinite space-like rod $[-\infty,z_1]$ in the $\psi$-direction
(thus $f_2(0,z)=0$ there),
a finite time-like rod $[z_1, z_2]$ ($f_0(0,z)=0$),
a second (and finite) space-like rod $[ z_2,z_3]$ in the $\psi$-direction, 
where $f_2(0,z)=0$ again, and 
a semi-infinite space-like rod $[z_3,\infty]$ ($f_3(0,z)=0$)
in the $\varphi$-direction (see Figure 1b).
The  metric functions $f_i$ of the static black ring\footnote{Note that the function $f_1(0,z)$ behaves as $1/|z-z_i|$ as
$z\to z_i$.} are given  by \cite{Emparan:2001wk},\cite{Harmark:2004rm}
\begin{eqnarray}
\nonumber
&&f_0 =\frac{R_2+\xi_2}{R_1+\xi_1},~~
f_1 = \frac{(R_1+\xi_1+R_2-\xi_2)((1-c)R_1+(1+c)R_2+2c R_3)}{8(1+c )R_1R_2R_3},
\\
\label{BR5d}
&&f_2 =\frac{(R_2-\xi_2)(R_3+\xi_3)}{R_1-\xi_1},~~f_3 =R_3-\xi_3~,
\end{eqnarray}
  where
\begin{eqnarray}
\label{rel1}
\xi_i=z-z_i,~~
R_i=\sqrt{r^2+\xi_i^2}
~~~~
{\rm and~}~~~~
z_1=-a,~~z_2=a,~~z_3= b,
\end{eqnarray}
$a$ and $ b$ being two positive constants, with $c=a/b<1$.
The mass, event horizon area  and Hawking temperature of the $d=5$ static black ring are:
\begin{eqnarray}
M^{(5)}=\frac{3aV_3}{4\pi G},~~
A_H^{(5)}=8a^2V_3\sqrt{\frac{2}{a+b}},~~
T^{(5)}=\frac{1}{4 \pi a}\sqrt{\frac{a+b}{2}},
\end{eqnarray}
(with $V_3=2 \pi^2$ the area of the three-sphere and $G$ the Newton constant). 
Although this solution is asymptotically flat\footnote{The $d=5$ static black ring solution in
  \cite{Emparan:2001wk} admits an alternative interpretation as 
  a ring sitting on the rim of a membrane that extends to infinity (this is found by requiring 
that the periodicity
of $\psi$ is $2\pi$ on the finite $\psi$-rod).
However, the asymptotic metric is a deficit membrane in this case.
}, it contains a conical excess angle $\delta$
for the  finite $\psi$-rod
\begin{eqnarray}
\delta=2\pi \left(1-\sqrt{\frac{b+a}{b-a}}\right).
\end{eqnarray}

\subsection{The Schwarzschild-Tangherlini black hole in $d\geq 5$ dimensions}
In principle, some $d>5$ black objects can also be constructed by using a similar construction
based on imposing a rod structure\footnote{This possibility was already mentioned in Ref. \cite{Kudoh:2006xd},
which, to our knowledge, is the only attempt in the literature to  numerically construct
asymptotically Minkowski solutions with a  nonspherical horizon topology.}.
The starting point is the observation that
the Minkowski spacetime is recovered within the framework (\ref{metric})
for
\begin{eqnarray}
\label{Mink}
 f_0=1,~~
 f_1=\frac{1}{2\sqrt{r^2+z^2}},~~
f_2= \sqrt{r^2+z^2}+z,~~~
f_3= \sqrt{r^2+z^2}-z,
\end{eqnarray}
(note that these expressions\footnote{The coordinate transformation
$r=\frac{1}{2}R^2 \sin 2 \theta,~~z=\frac{1}{2}R^2 \cos 2 \theta$
(with $0\leq R<\infty$, $0\leq \theta  \leq \pi/2$), leads to a more common form of the flat spacetime metric:
$ ds^2=-dt^2+dR^2+R^2 (d\theta^2+\cos^2 \theta d\psi^2+\sin^2 \theta d\Omega^2_{d-4}).$
} do not depend on the spacetime dimension $d$).
One can see that $f_2$ vanishes for 
$r=0, z<0$, and $f_3$ for $r=0, z>0$,
which, in the language of the Weyl formalism, 
 corresponds to two semi-infinite  rods $ [-\infty,0]$ and $ [0,\infty]$.

\begin{figure}[ht]
\hbox to\linewidth{\hss%
	\resizebox{6cm}{4cm}{\includegraphics{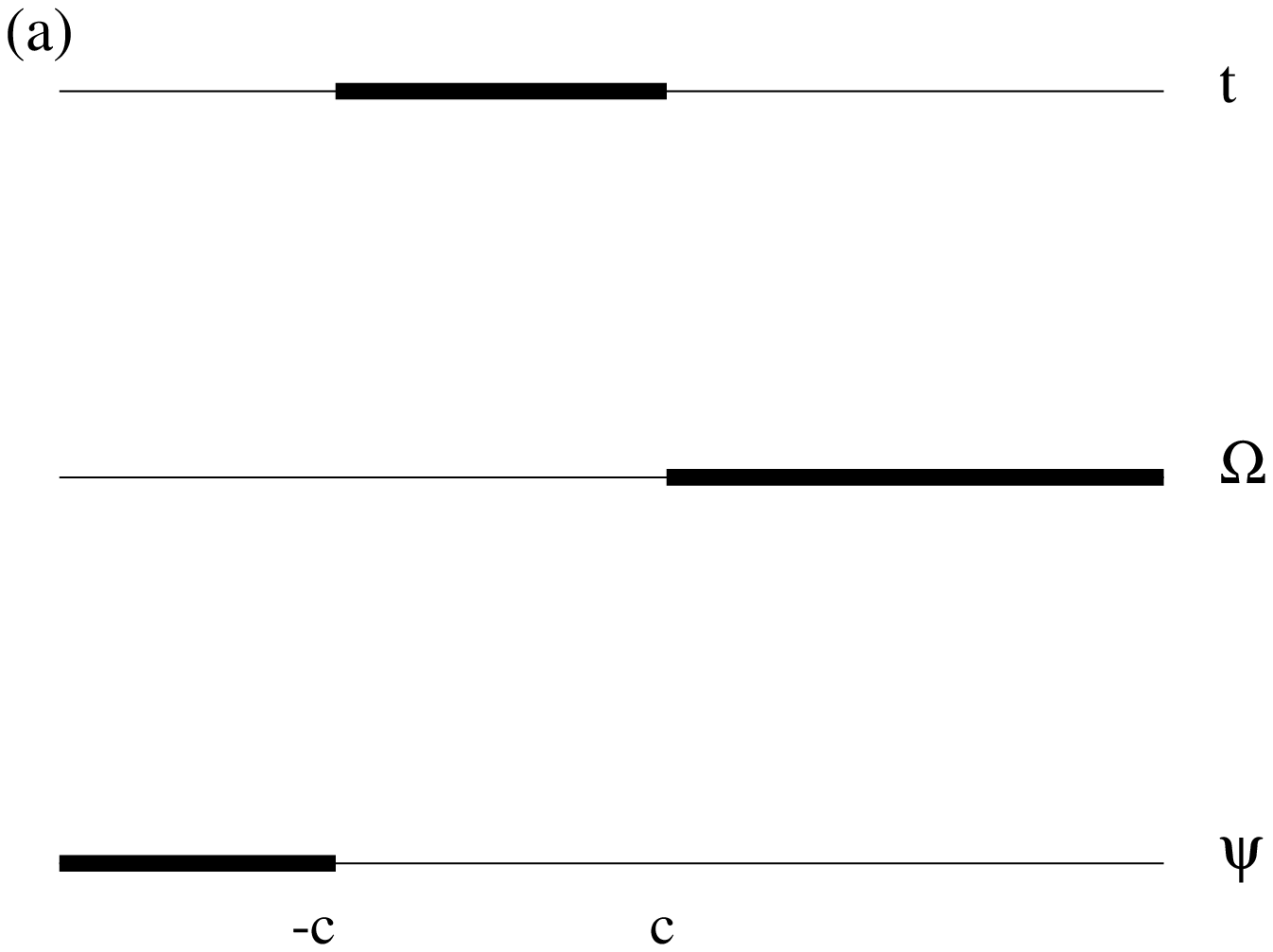}}
\hspace{15mm}%
        \resizebox{6cm}{4cm}{\includegraphics{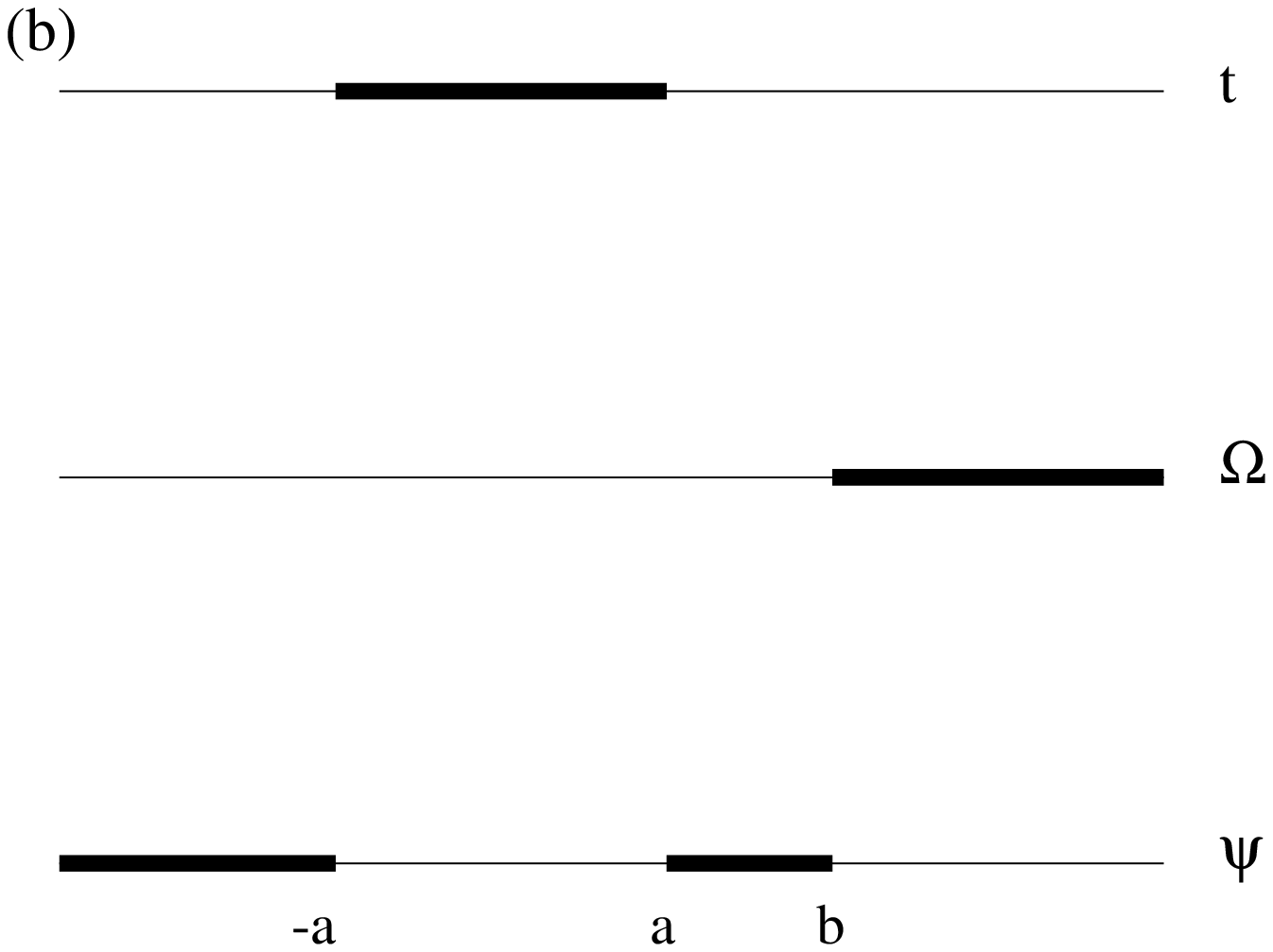}}	
\hss}
\vspace{0.4cm}
\caption{\small  The  rod structure of the solutions is shown for the Schwarzschild-Tangherlini
solution (left) and the black hole with $S^2\times S^{d-4}$ topology
of the horizon (right).
The thin lines denote the $z-$axis and the thick lines denote the rods.
$\Omega$ stands for all directions on the $S^{d-4}$ sphere.
 }
\label{fig1}
\end{figure} 

Moreover, black holes with a spherical event horizon can also be written 
within the ansatz (\ref{metric}) as proven by the $d\geq 5$
Schwarzschild-Tangherlini solution.
A straightforward but cumbersome  computation leads to the following expression for the
metric functions in this case 
\begin{eqnarray}
\label{metric-ST}
&&f_0=\left(\frac{v^{(d-3)/2}-1}{v^{(d-3)/2}+1}\right)^2,~~~
 f_1=c \frac{(v^{(d-3)/2}+1)^{4/(d-3)}}
 {4v(z^2\frac{(v^2-1)^2}{(v^2+1)^2}+r^2\frac{(v^2+1)^2}{(v^2-1)^2})},
 \\
 \nonumber
&&f_2=\frac{1}{2v}\frac{(v^{(d-3)/2}+1)^{4/(d-3)}}{v^2+1}(c(v^2+1)+2zv),~~
  f_3=\frac{1}{2v}\frac{(v^{(d-3)/2}+1)^{4/(d-3)}}{v^2+1}(c(v^2+1)-2zv),
\end{eqnarray}
where
\begin{eqnarray}
\nonumber
&&v=\frac{1}{c}\left(r^2+z^2+{\cal P}+\sqrt{2}
\sqrt{(r^2+z^2)^2+c^2(r^2-z^2)+(r^2+z^2){\cal P}}\right )^{1/2},
\\
\nonumber
&&~~{\rm and~~}
{\cal P}=\sqrt{c^4+2c^2(r^2-z^2)+(r^2+z^2)^2 }, 
\end{eqnarray}
with $c$ an arbitrary positive constant\footnote{One can show that 
for $d=5$ these expressions reduce to those in \cite{Emparan:2001wk}.}. 

For any $d>4$,
the rod structure of the Schwarzschild-Tangherlini black hole
consists of 
a semi-infinite space-like rod $[-\infty,-c]$
(with $f_2(0,z))=0$ there),
a finite time-like rod $[-c, c]$ ($f_0(0,z)=0$) and a  
semi-infinite space-like rod $[c,\infty]$ (with a vanishing $f_3(0,z)$)
in the $\Omega$-direction (see Figure 1a).
The basic properties of the  Schwarzschild-Tangherlini
black hole can easily be rederived within this metric ansatz.
Again, requiring the absence of a conical singularity imposes a 
periodicity $2 \pi$
for the coordinate $\psi$. 

\section{Black holes with $S^2\times S^{d-4}$ topology of the event 
horizon}
It is natural to conjecture that the ansatz (\ref{metric})
can be used to construct $d>5$ solutions with a nonspherical horizon topology.
Perhaps the simplest case is provided by black holes with 
a topology of the horizon which is  $S^2\times S^{d-4}$.
The rod structure of these solutions is similar to the case of 
$d=5$ static black rings, with the $\varphi-$angle there
replaced by the $\Omega_{d-4}$ sphere, as shown in Figure 1b.

\subsection{The rod structure and boundary conditions}
The new solutions in this paper have first a 
semi-infinite space-like rod $[-\infty,-a]$
with the following expansion\footnote{In all relations below, the functions $f_{ik}(z)$ are 
solutions of a complicated set of nonlinear second order differential equations.} of the metric function as $r\to 0$
(except for $z\to \pm a,b$):
\begin{eqnarray}
\label{rod1}
&&f_0(r,z)=f_{00}(z)+r^2f_{02}(z)+\dots,~~f_1(r,z)=f_{10}(z)+r^2f_{12}(z)+\dots,
\\
\nonumber
&&f_2(r,z)= r^2f_{22}(z)+r^4f_{24}(z)+\dots,~~f_3(r,z)=f_{30}(z)+r^2f_{32}(z)+\dots,
\end{eqnarray}
the constraint equation $G_r^z=0$ implying  $\lim_{r\to 0}r^2 f_1/f_2=c_1$.
Next, there is a finite time-like rod $[-a, a]$ corresponding to the horizon, with
\begin{eqnarray}
\label{rod2}
&&f_0(r,z)=r^2f_{02}(z)+r^4f_{04}(z)+\dots,~~f_1(r,z)=f_{10}(z)+r^2f_{12}(z)+\dots,
\\
\nonumber
&&f_2(r,z)= f_{20}(z)+r^2f_{22}(z)+\dots,~~f_3(r,z)=f_{30}(z)+r^2f_{32}(z)+\dots,
\end{eqnarray}
(with $\lim_{r\to 0}r^2 f_1/f_0=const.$),
and a finite space-like rod $[a,b]$
along the $\psi$-direction, where  the metric functions
have an expression similar to (\ref{rod1}); the constraint equation $G_r^z=0$ implies there  
$\lim_{r\to 0}r^2 f_1/f_2=c_2$, where $c_1,c_2$ are arbitrary positive constants.

The case of the $\Omega-$rod ($i.e.$  $[b,\infty]$) is more involved.
The expansion here is  
\begin{eqnarray}
\label{rod3}
&&f_0(r,z)=f_{00}(z)+r^2f_{22}(z)+\dots,~~f_1(r,z)=f_{32}(z)+r^2f_{12}(z)+\dots,
\\
\nonumber
&&f_2(r,z)=f_{20}(z)+r^2f_{22}(z)+\dots,~~f_3(r,z)= r^2f_{32}(z)+r^4f_{34}(z)+\dots,
\end{eqnarray}
Here the relation $\lim_{r\to 0}r^2f_1/f_3=1$ follows, which is a feature of the $d>5$ case.
In five spacetime dimensions the value of this limit is not fixed.

The obvious boundary conditions for large $r,z$ is that $f_i$ approach the Minkowski background
functions (\ref{Mink}).
In practice, we 
have found it convenient to take
\begin{eqnarray}
\label{ans1}
f_i=f_i^{0}F_i ,
\end{eqnarray}
where $f_i^{0}$ are some background functions, given by the metric functions of the $d=5$ static
black ring solution (\ref{BR5d}). 
The advantage of this approach is that $f_i$ will automatically satisfy the
desired rod structure.

The equations satisfied by $F_i$ can easily be derived from (\ref{eqs1}).
As for the boundary conditions, the relations (\ref{rod1})-(\ref{rod3})
together with the expressions (\ref{BR5d}) of the background functions $f_i^{0}$
imply
\begin{eqnarray}
\label{bc-psi}
\nonumber
&&\partial_r F_i|_{r=0}=0,~~~~{\rm for~~}-\infty<z\leq b,
\\
\label{bc-Omega}
&&\partial_r F_0|_{r=0}=0,~~\partial_r F_1|_{r=0}=0,~~
\partial_r F_2|_{r=0}=0,~~F_1|_{r=0}=F_3|_{r=0}~~~~{\rm for~~}b<z\leq \infty ,
\end{eqnarray}
and $F_i=1$ as $r\to \infty$ or $z\to \pm \infty$.

The constraint equation $G_r^z=0$ implies 
 $F_2/F_1=const.$ on the $\psi$-rods.
Now, to be consistent with the assumption of asymptotic flatness, one finds $const.=1$
for $-\infty<z\leq -a$. The value of this ratio for the finite rod with $a<z\leq b$
results from numerics.

\subsection{Global quantities}
The horizon metric is given by
\begin{eqnarray}
d\sigma^2=f_1(0,z)dz^2+f_2(0,z)d\psi^2+f_3(0,z)d\Omega_{d-4}^2,
\end{eqnarray}
with  $-a \leq z\leq a$.
Since the orbits of $\psi$ shrink  to zero at $-a$ and $a$ while 
the area of $S^{d-4}$ does not vanish anywhere, the topology of the horizon 
is $S^2\times S^{d-4}$.
 
The
event horizon area is given by
\begin{eqnarray}
A_H=\Delta \psi V_{d-4} \int_{-a}^a dz \sqrt{f_1 f_2f_3^{d-4}}=
2 \Delta \psi V_{d-4}  \frac{2^{(d-4)/2}a}{\sqrt{a+b}}  \int_{-a}^a dz~(b-z)^{(d-5)/2} \sqrt{F_1 F_2F_3^{d-4}},
\end{eqnarray}
where $V_{d-4}$ is the area of the unit sphere $S^{d-4}$ and $\Delta \psi$ the periodicity of the 
angular coordinate $\psi$ on the horizon.

%
\begin{figure}[ht]
\hbox to\linewidth{\hss%
	\resizebox{8cm}{6cm}{\includegraphics{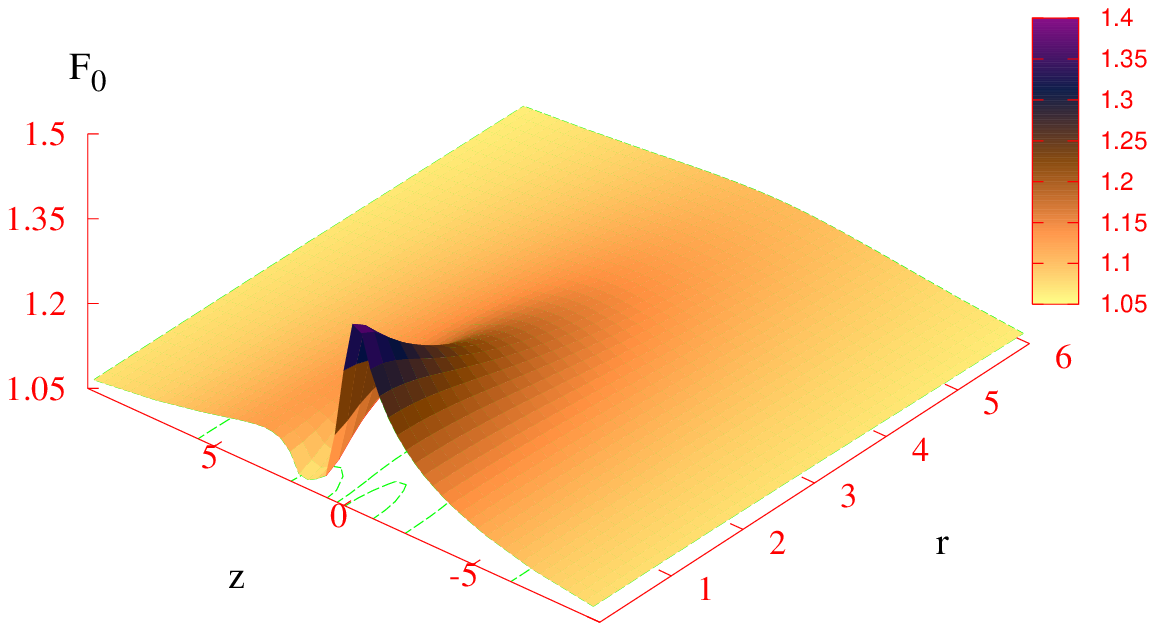}}
\hspace{5mm}%
        \resizebox{8cm}{6cm}{\includegraphics{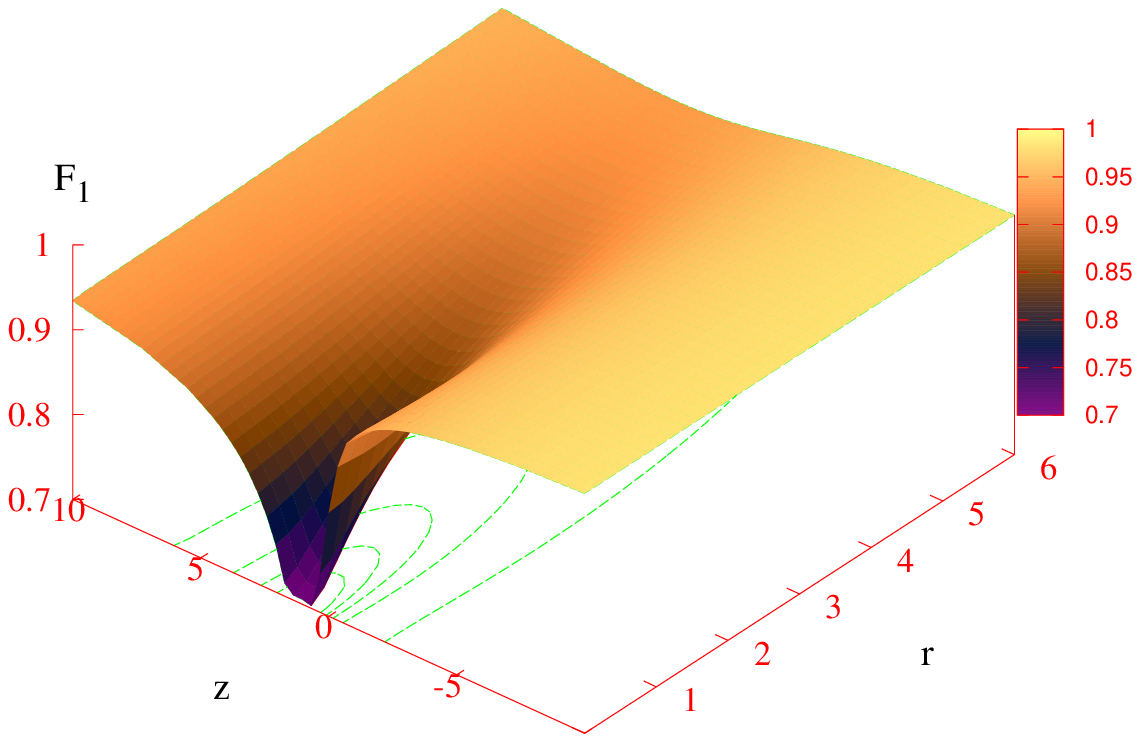}}	
\hss}
\end{figure}
\begin{figure}[ht]
\hbox to\linewidth{\hss%
	\resizebox{8cm}{6cm}{\includegraphics{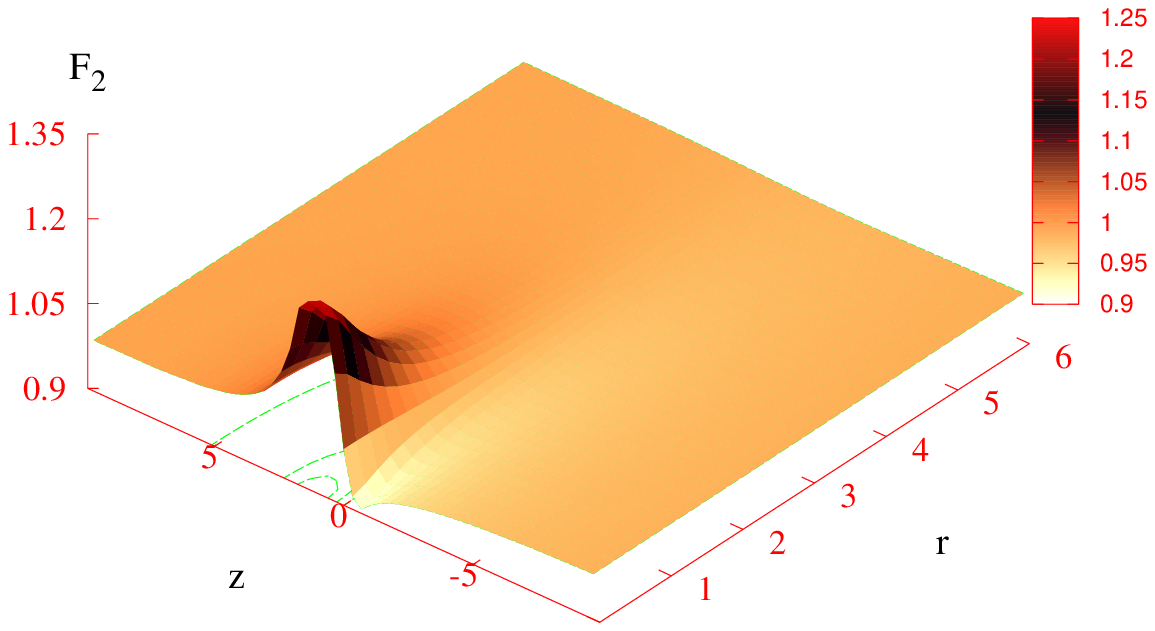}}
\hspace{5mm}%
        \resizebox{8cm}{6cm}{\includegraphics{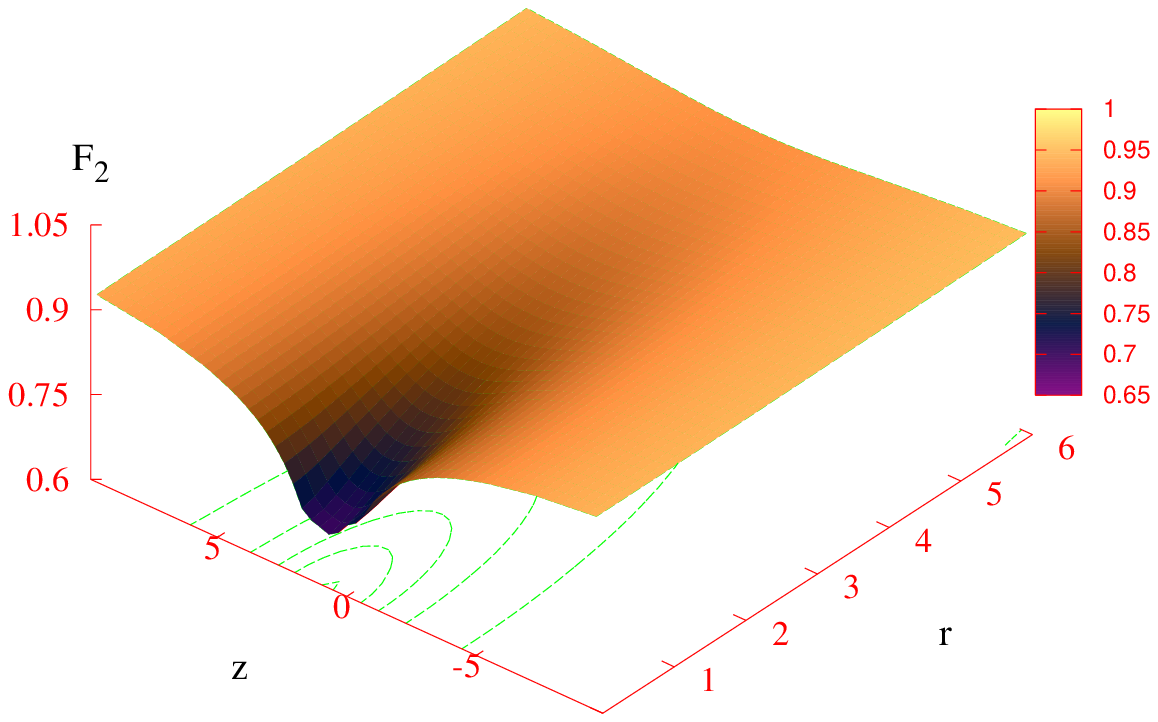}}	
\hss}\caption{\small  The profiles of the metric functions $F_i$ for a typical
$d=6$ solution with $a=0.5$, $b=1.7$.
 }
\label{fig3d}
\end{figure}
The Hawking temperature can be computed from the surface gravity or by requiring regularity 
on the Euclidean section
\begin{eqnarray}
T=\frac{1}{4 \pi a}\sqrt{\frac{a+b}{2}}\sqrt{\frac{F_0}{F_1}}~,
\end{eqnarray}
where the constraint equation $G_r^z=0$ guarantees that the ratio $F_0/F_1$ is constant on the event horizon.
At infinity, the Minkowski background is approached, with $\Delta \psi=2 \pi$ there.
The mass $M$ of the solutions can be read from the asymptotic
expression for $f_0$
\begin{eqnarray}
\label{gtt}
f_0\sim 1-\frac{16 \pi G M}{(d-2)V_{d-2}(r^2+z^2)^{(d-3)/4}}+\dots~.
\end{eqnarray}
The solutions satisfy also the Smarr law
\begin{eqnarray}
\label{smarr}
M=\frac{d-2}{4G(d-3)}T A_H~.
\end{eqnarray}

All solutions we have found have $F_2/F_1\neq 1+2a/(b-a)$
for the finite $\psi-$rod.
Thus, for  $a\leq z \leq b$, the $(z,\psi)$ part of the metric
describes a surface that is topologically $S^2$ with a conical singularity at one of the poles.
This conical singularity prevents this black object to collapse 
to form a spherical black hole horizon.
The value of the conical excess of $\psi$ 
for $a<z < b$ is 
\begin{eqnarray}
\delta=2\pi \left(1-\sqrt{\frac{b+a}{b-a}}\sqrt{\frac{F_2}{F_1}} \right).
\end{eqnarray}
We have found it convenient to introduce the quantity
\begin{eqnarray}
\label{rel-delta}
\bar \delta=\frac{\delta/(2\pi)}{1-\delta/(2\pi)},
\end{eqnarray}
which has a finite range, $-1\leq \bar \delta\leq 0$, and measures the 'relative angular excess'.

\begin{figure}[ht]
\hbox to\linewidth{\hss%
	\resizebox{8cm}{6cm}{\includegraphics{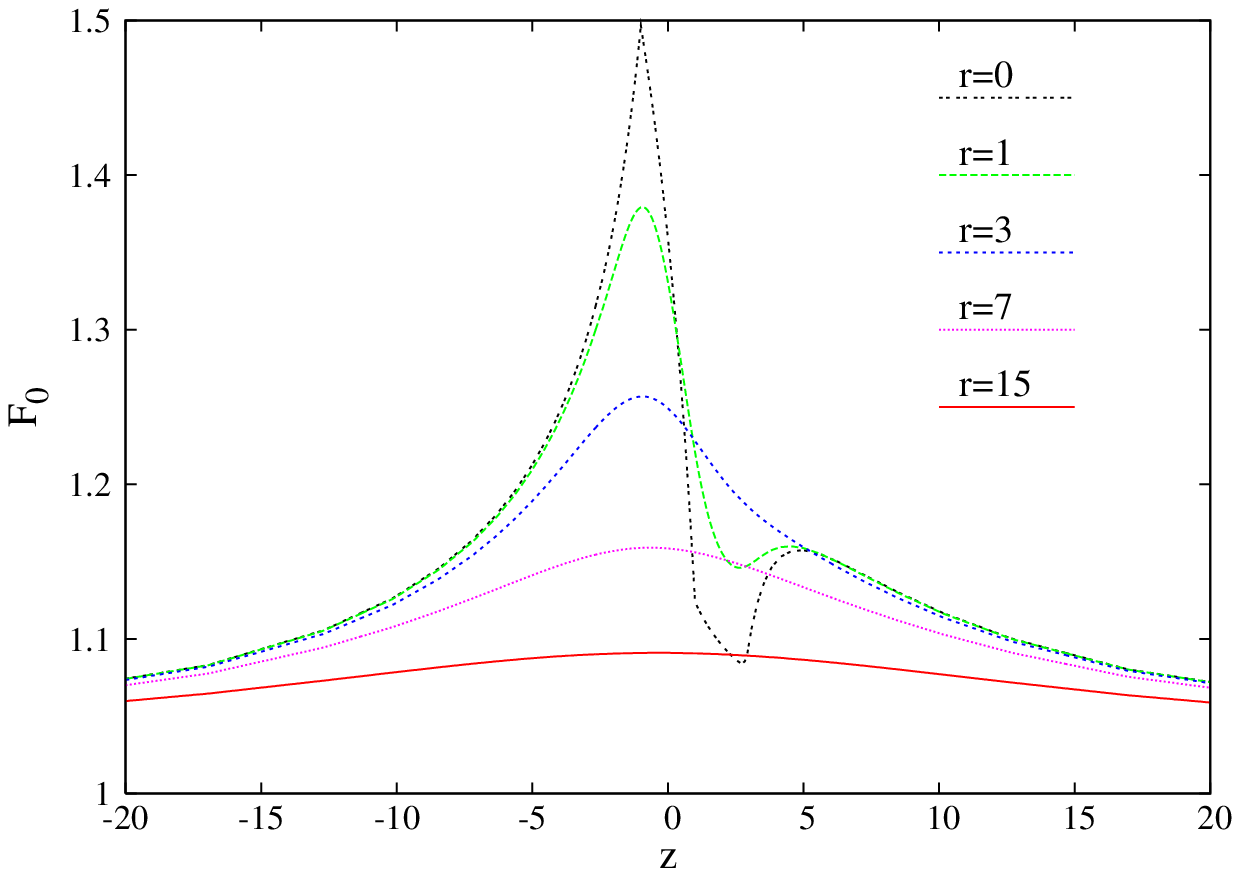}}
\hspace{5mm}%
        \resizebox{8cm}{6cm}{\includegraphics{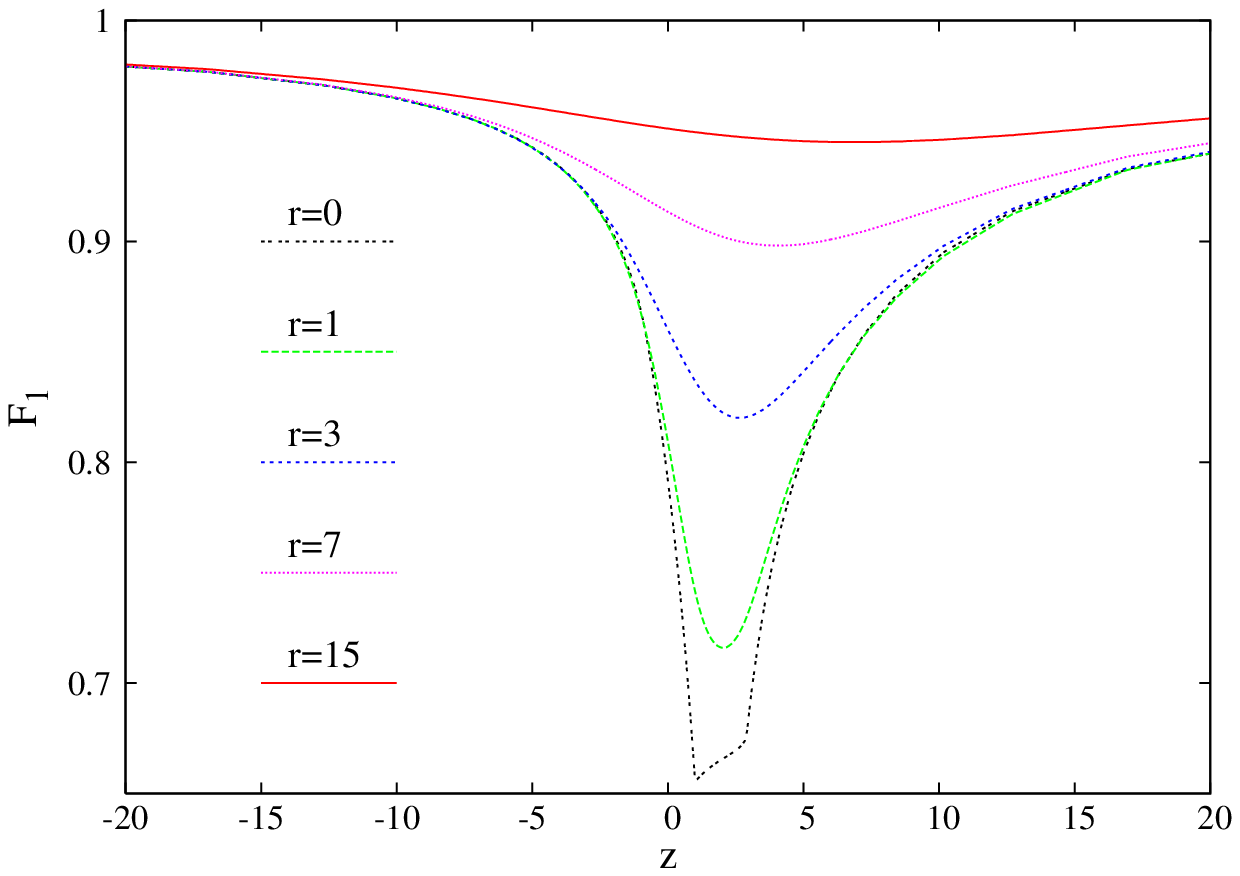}}	
\hss}
\end{figure}
\begin{figure}[ht]
\hbox to\linewidth{\hss%
	\resizebox{8cm}{6cm}{\includegraphics{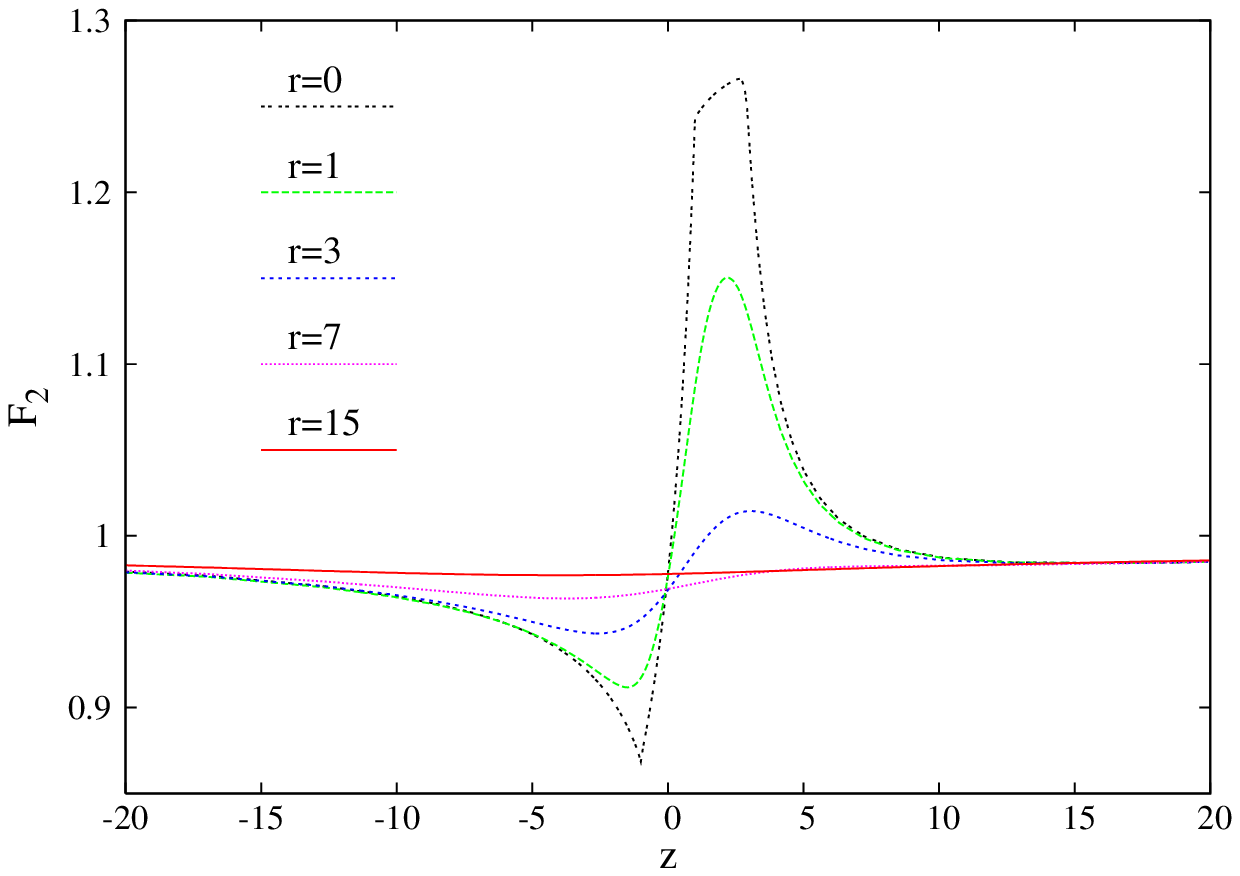}}
\hspace{5mm}%
        \resizebox{8cm}{6cm}{\includegraphics{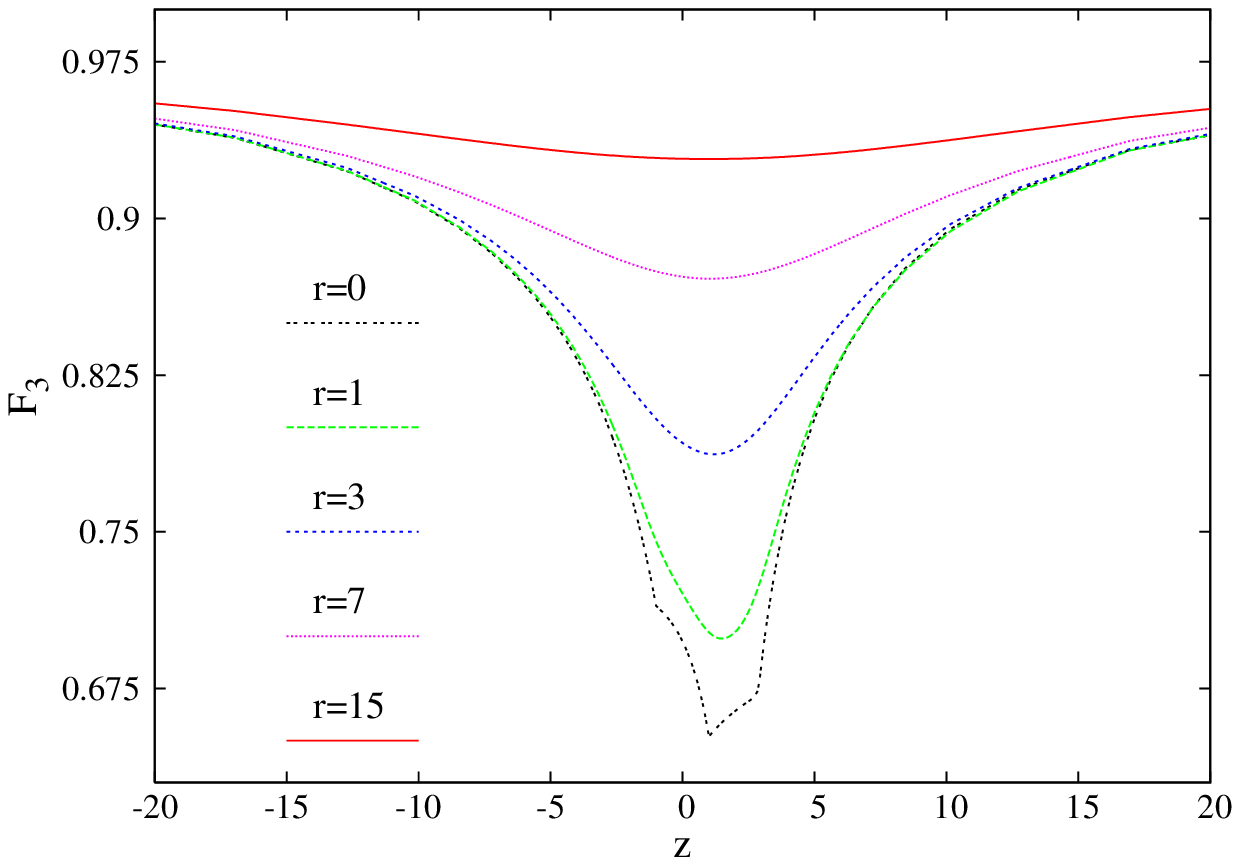}}	
\hss}\caption{\small  The metric functions $F_i$ 
are shown as a function of $z$ for several values of $r$.
The relevant parameters here are
 $d=6$, $a=1$, $b=2.85$.
 }
\label{fig3}
\end{figure}

\section{$d=6,7$ numerical solutions}
We solve the resulting set of four coupled non-linear
elliptic partial differential equations numerically,
subject to the above boundary conditions.
All numerical calculations  
are performed by using the program FIDISOL/CADSOL \cite{schoen},
which uses a  Newton-Raphson method.

First, one introduces  the new compactified coordinates 
$x=r/(1+r)$, $u=\arctan(z)$. 
The equations for $F_i$ are then discretized on a non-equidistant grid in
$x$ and $u$. 
Typical grids used have sizes $80 \times 160$,
covering the integration region
$0\leq x \leq 1$ and $-\pi/2\leq u \leq \pi/2$. 
(See \cite{schoen} and \cite{kk}
for further details and examples for the numerical procedure.) 
In this scheme, the input parameters are the positions of the rods fixed by $a$ and $b$
and the value $d$ of the spacetime dimension.
To obtain  solutions with $S^2\times S^{d-4}$ horizon topology, 
one starts with the $d=5$ solution 
 as initial guess ($i.e.$ $F_i=1$) and increases the value of $d$ slowly.
The iterations converge, and, in principle, repeating the procedure one obtains
in this way solutions for arbitrary $d$.
The physical values of $d$ are integers.
We have studied   solutions in $d=6,7$ dimensions in a systematic way. 
Solutions with $d>7$ are also likely to exist;
however, their study may require a different numerical method.

\begin{figure}[ht]
\hbox to\linewidth{\hss%
	\resizebox{8cm}{6cm}{\includegraphics{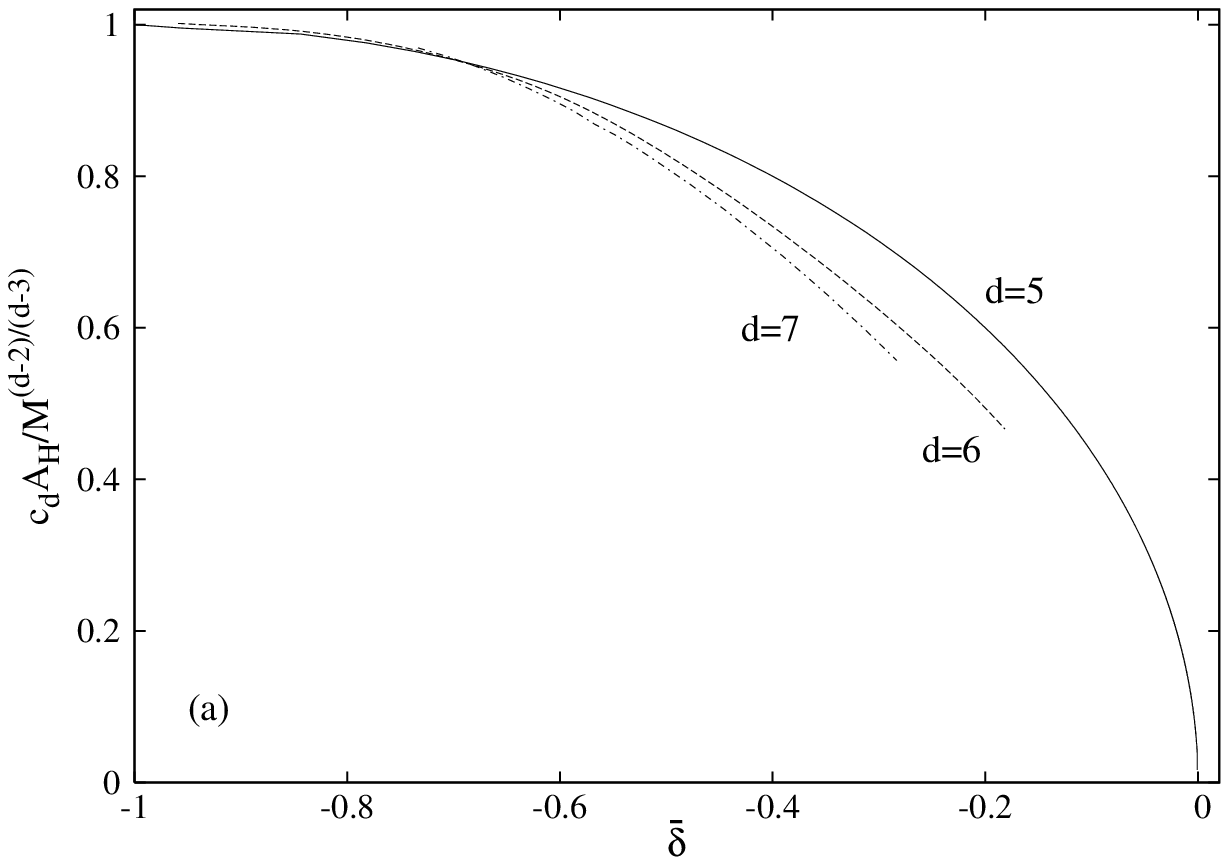}}
\hspace{5mm}%
        \resizebox{8cm}{6cm}{\includegraphics{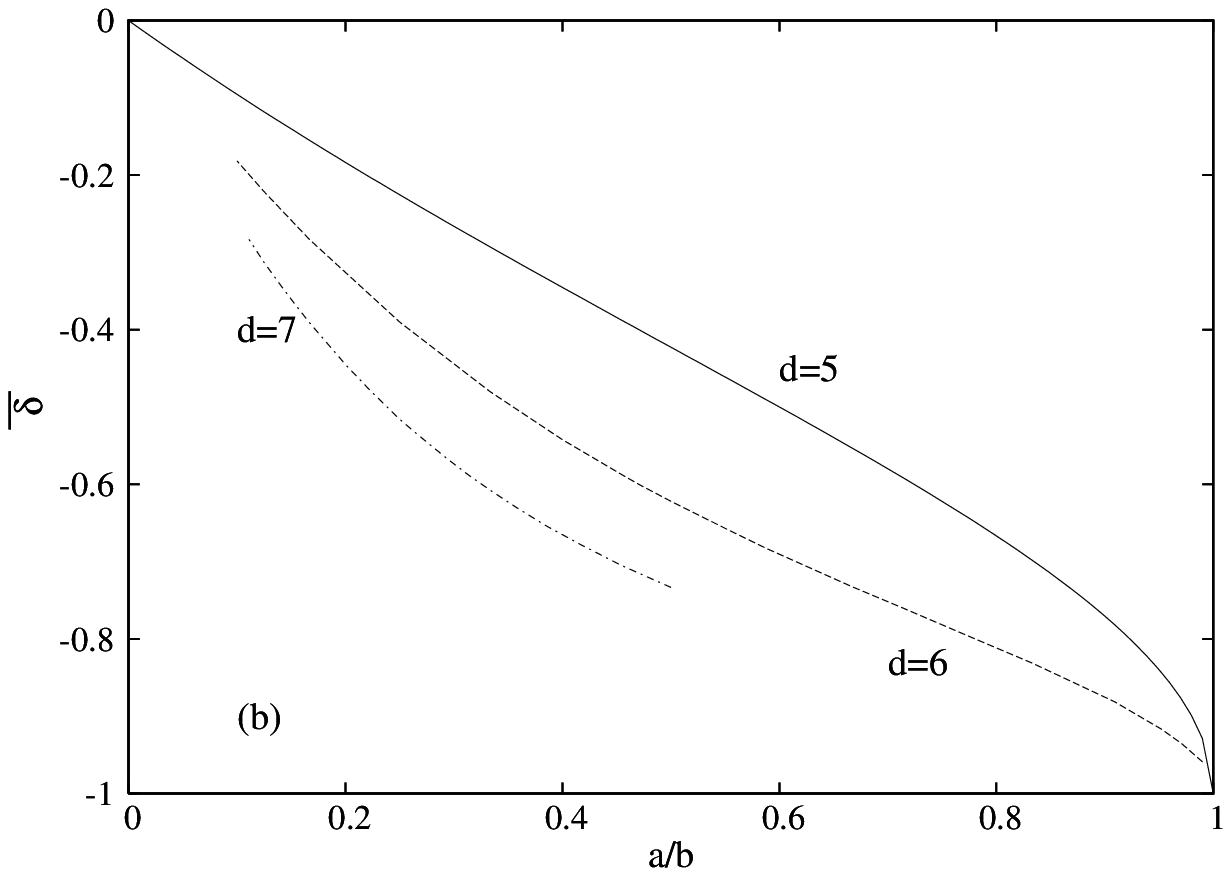}}	
\hss}\caption{\small {  (a):} The scale free ratio $A_H/M^{(d-2)/(d-3)}$
is shown as a function of the relative angular excess $\bar \delta$ 
as defined by the relation (\ref{rel-delta}). 
The value of the parameter $c_d$ there is $c_d=(d-2)/(16\pi)^{(d-2)/(d-3)}V_{d-2}^{1/(d-3)}$
and has been choosen such that the point $(1,-1)$ on the plot corresponds to the
Schwarzschild-Tangerlini black hole.
{ (b):} The relative angular excess $\bar \delta$  is shown as a function of the ratio 
between the two length scales $a/b$.
 }
\label{fig4}
\end{figure}

The problem has two length scales $a$ and $b$, roughly corresponding to the 
event horizon radius and the radius of the round $\Omega$-sphere.
It turns out that the above approach works  well in the intermediate region of the ratio $a/b$,
and fails to provide solutions with good accuracy if the two scales are widely separated or very close to each other
($i.e.$ $a/b\to 0$ or $a/b \to 1$). 
FIDISOL/CADSOL provides also an error estimate for each unknown function.
For $d=6$, 
the typical  numerical error 
for the functions is estimated to be lower than $10^{-3}$.

The Hawking temperature, event horizon area and conical excess are
encoded in the values of
the  functions $F_i, f_i^b$ at $r=0$.
The mass parameter $M$ is computed from the asymptotic 
form (\ref{gtt}) of the metric function $g_{tt}=-f_0$, the Smarr relation (\ref{smarr})
being used to verify the accuracy of the solutions. 

The functions $F_i$
change smoothly with the two  parameters $a$ and $b$.
Typical profiles of the solutions are presented  
 in Figures 2, 3.
One can see that the functions $F_i$ are smooth outside of the $z-$axis and show no sign of a singular behaviour.
The crucial point here is that the divergent behaviour of $f_i$ was already subtracted by the background
functions $f_i^{b}$.
We have verified that the Kretschmann scalar stays finite everywhere, in particular at $r=0$.

By fixing the position of the horizon and varying the length of the finite $\psi-$rod,
we have generated 
a branch of $d=6,7$ black objects with $S^2\times S^{d-4}$ horizon topology.
The picture we have found is very similar to that valid for $d=5$ static black rings.
First, all solutions have a conical excess $\delta$ on the finite $\psi-$rod.
Moreover, on the horizon, the area of the round $S^{d-4}$-sphere  is maximum for $z=-a$ and minimum at 
$z=a$, which shows that $\Delta \psi=2\pi$ there.  

In terms of the quantity $\bar \delta$ as defined in (\ref{rel-delta}), 
these black objects smoothly interpolate between two limits
(although these regions of the parameter space are difficult to approach numerically).
First,
as $a/b \to 1$, one finds $\bar \delta\to -1$ ($i.e.$ the conical excess $\delta\to -\infty$) and
 the Schwarzschild-Tangherlini metric is approached (the finite $\psi-$rod vanishes).
As the second $\psi-$rod extends to infinity ($a/b\to 0$), the 
radius  on the horizon of the round $S^{d-4}$-sphere increases and asymptotically it
becomes a $(d-4)$-plane, while  $\bar \delta\to 0$.
Here one
expects to recover, after a suitable rescaling,  the 
four dimensional Schwarzschild black hole uplifted to $d$ dimensions\footnote{For the
static $d=5$ black ring solution, this limit is found by taking 
$r\to \sqrt{2b}\bar r$,
$z\to \sqrt{2b}\bar z$,
$\varphi\to \bar \varphi/\sqrt{2b}$,
followed by $b\to \infty$.
This results in the Schwarzschild black string solution.
}  ($i.e.$ a black $(d-4)$-brane).
These features are illustrated in Figure 4.
For completeness, we have shown also the corresponding data for 
the $d=5$ static black rings.

\section{Further remarks}
 In the absence of analytical methods to construct $d>5$ black hole solutions with nonspherical horizon 
 topology, 
a numerical approach of this problem seems to be a reasonable task.
In this work we have presented such a construction for $d=6,7$ static black objects 
with $S^2\times S^{d-4}$ topology of the horizon, as a first step before 
approaching more complex situations.
The existence of the $d=6$ solution is not a surprise, since the results in \cite{Helfgott:2005jn}
show that this topology 
is one of the few allowed for black holes in six dimensions. 
However, given the presence of a conical excess angle, the  solutions we have found
are presumably unstable and their physical relevance is obscure.

The technique  proposed in this paper can easily be extended for  other types of $d>5$ static   
black objects ($e.g.$ a superpositions of a Schwarzschild-Tangerlini black hole
and a configurations with a $S^2\times S^{d-4}$ topology of the horizon).
Also, in principle, the inclusion of rotation in this scheme is straightforward.
For example, similar to the $d=5$ case, the conical singularity of the solutions with $S^2\times S^{d-4}$
topology of the horizon
could presumably be eliminated if the $\Omega$-sphere would rotate. 
However, this leads to a difficult numerical problem, since the equations would depend 
on at least three variables.
Another possible direction would be to consider more complicated versions of the 
static ansatz (\ref{metric}) in $d>6$ ($e.g.$ replace the angular direction $\psi$ with a sphere $S^n$). 
In the absence of rotation,  one would expect such vacuum solutions to possess again
some unphysical features.

Moreover, the inclusion of some matter fields is unlikely to cure the conical singularity.
This is the case for the Einstein-Maxwell-dilaton generalization of the  solutions here, which extremize the action
\begin{eqnarray}
\label{actionEMd}
I=\frac{1}{16 \pi G}\int d^dx \sqrt{-g} \left(R-
\frac{1}{2} g^{\mu \nu} \partial_\mu \phi \partial_\nu \phi
-\frac{1}{4} e^{-2\bar a \phi} F^2  \right),
\end{eqnarray}
where $F=dA$.
Starting with a vacuum configuration (\ref{metric}), a Harrison transformation 
(see $e.g.$ \cite{Gal'tsov:1998yu})
leads to an EMd solution with line element
\begin{eqnarray} 
\label{forma2}
&ds^2~=-f_0(\cosh^2 \beta-\sinh^2 \beta f_0)^{-2\alpha (d-3)}dt^2+
(\cosh^2 \beta-\sinh^2 \beta f_0)^{2\alpha}(f_1(dr^2+dz^2)
+f_2d\psi^2+f_3d\Omega_{d-4}^2),~{~}~
\end{eqnarray}
and matter fields 
\begin{eqnarray} 
\label{forma3}
A_\mu=\sqrt{2(d-2)\alpha} 
\frac{\tanh \beta~e^{\bar a\phi_0}f_0}
{\cosh^2 \beta-\sinh^2 \beta f_0 }\delta_{\mu t},~~
\phi=\phi_0-2\bar a(D-2)\alpha\log(\cosh^2 \beta-\sinh^2 \beta f_0), 
\end{eqnarray}
where $\beta,~\phi_0$ are arbitrary real constants
and 
$\alpha=(2\bar a^2(d-2)+d-3)^{-1}.$
Note that the Harrison transformation does not affect the rod structure
of the metric. Thus, one finds in this way charged black objects with  $S^2\times S^{d-4}$
topology of the horizon.
For $d=5$, this is the solution presented in \cite{Kunduri:2004da} 
by using a different coordinate system. 

Unfortunately, one can easily see that for both $d=5$ and $d=6,7$, 
the line element (\ref{forma2}) has the same conical singularity as
the vacuum seed solution.
In principle, this singularity can be removed 
by "immersing" the solutions in a background gauge field, which
 requires to apply a second Harrison transformation 
(note that the resulting solutions would not be
asymptotically flat; for $d=5$, such a construction has been presented in \cite{Kunduri:2004da}).
The seed solution here is 
(\ref{forma2}), (\ref{forma3}), with the electric field dualized, $\tilde F =e^{-2 a \phi}\star F$ (with $\tilde F=dB$), 
and $\hat \phi =-\phi$.
Then, the conical singularity vanishes for a critical value of the parameter in the second Harrison
transformation.
However, this last point requires  knowledge of the explicit form of the magnetic potential
$B$ on the $\psi$-rods, which seems not possible for the $d>5$ numerical solutions.

We close this work by remarking that, on general grounds, a numerical  approach
works if the length scales involved are not widely separated,
which is just the opposite of the approximate construction in \cite{Emparan:2007wm},\cite{Emparan:2009cs}.
Thus these methods are complementary. 
It would be interesting to construct solutions with
 $S^{d-3}\times S^1$ topology of the event horizon, which were already considered in \cite{Emparan:2007wm}.
Supposing that the static limit of these  black rings  can be constructed by using the ansatz  (\ref{metric}),
their rod structure can be read from Figure 1b by  interchanging the $\psi$ and $\Omega$-rods there
($e.g.$ there is only one $\psi-$rod for  $[b,\infty]$, etc.).
The boundary conditions in this case are still given by (\ref{rod1})-(\ref{rod2}).
An unexpected feature here is that these configurations would
possess no conical singularities. 
This difference to the static $d=5$ black rings originates in the presence of  $(d-5)$ factors in the field equations (\ref{eqs1})).
This counter-intuitive result sheds doubt on the possibility to construct $d>5$ static black rings
with a regular horizon within the ansatz 
(\ref{metric}) with a dependence on only two coordinates. 
\\
\\
{\bf\large Acknowledgements} 
\\
B.K. gratefully acknowledges support by the DFG. 
The work of E.R. was supported by a fellowship from the Alexander von Humboldt Foundation. 
E.R. would like to thank Cristian Stelea for interesting remarks on a draft of this paper.

 \begin{small}
 
 \end{small}

 \end{document}